\documentclass[review]{elsarticle}

\usepackage{lineno,hyperref,float}
\modulolinenumbers[5]

\journal{Journal of \LaTeX\ Templates}

\begin{document}

\begin{frontmatter}

\title{{\it TriSol}: a major upgrade of the {\it TwinSol} RNB facility}

\author{P.D. O'Malley}

\author {T. Ahn, D.W. Bardayan, M. Brodeur, S. Coil }

\author {J.J. Kolata\fnref{myfootnote2}\corref{mycorrespondingauthor}}
\cortext[mycorrespondingauthor]{Corresponding author}
\ead{jkolata@nd.edu}
\fntext[myfootnote2]{Emeritus professor}

\address{Dept. of Physics and Astronomy, University of Notre Dame, Notre Dame, IN 46556, United States}


\begin{abstract}
We report here on the recent upgrade of the {\it TwinSol} radioactive nuclear beam (RNB) facility at the University
of Notre Dame.  The new {\it TriSol} ~system includes a magnetic dipole to provide a second beamline
and a third solenoid which acts to reduce the size of the radioactive beam on target. 
\end{abstract}


\end{frontmatter}

%\linenumbers

\section{Introduction }

The first radioactive nuclear beam setup \cite{RIB1} at the University of Notre Dame, used from 1987-1995, was replaced by the
{\it TwinSol} facility \cite{TwinSol1,TwinSol2,TwinSol3,TwinSol4, TwinSol5, TwinSol6} which became operational in 1998.  Both of these were the result of a collaboration between the University of Notre Dame and a group at the University of Michigan headed by Prof. F. D. Becchetti.  
\begin{figure}
[!htb]
\center
\includegraphics*[width=0.99\columnwidth]{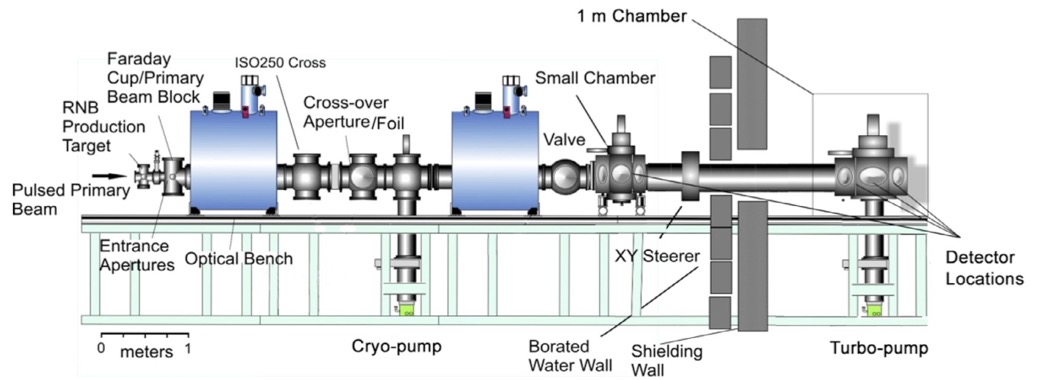}
\caption{\label{twinsol}(Color online) {The \it TwinSol} radioactive nuclear beam facility. Figure from Ref. \cite{TwinSol6}.}
\label{fig:1} 
\end{figure}
\newpage
\noindent
{\it TwinSol} (Fig.\ref{fig:1}) consists 
of a pair of 30 cm
bore, 6T superconducting solenoids contained within low-loss cryostats ($< $0.1 LHe/h) having a holding time
of $>$2 months.  The solenoids are air-core (hence no iron yokes) and are operated in persistent mode.  As a result, their 
magnetic fields do not suffer from hysteresis and scale exactly with the applied currents.  However, the fields extend to a large 
distance (up to 0.1 T at 2 m on axis at maximum field) and this must be accounted for both in computing the required magnet current and in safety concerns for those working  near the magnets when they are energized.
\\
\\
Referring to Fig.\ref{fig:1} above, a variable aperture (0-30 mm dia.) at the crossover point between the two solenoids provides momentum analysis
and therefore a degree of isotope separation.  In some cases, an absorber foil was also located there to provide further separation via differential energy loss.  Most of the early experiments were carried out in the small scattering chamber close to the second solenoid.  
In this location, the beam spot size was typically 5-6 mm full width at half maximum (FWHM).  However, in later experiments involving
neutron or $\gamma$-ray detection, a shielding wall consisting of borated water followed by concrete blocks was inserted as shown. Experiments were also carried out in the larger scattering chamber illustrated, or in other apparatus such as the prototype active-target
time projection chamber (pAT-TPC) \cite{pAT-TPC} from Michigan State University (MSU) or the neutron array described in \cite{narray}.
\\
\\
The primary target was initially situated within an ISO-100 cross, followed by an ISO-200 cross containing an entrance aperture and
a Faraday cup having an outer diameter of 2.5 cm.  Typically this setup accepted particles scattered between 3$^\circ$ and 6$^\circ$
 in the laboratory (lab) frame, although larger collimators accepting up to 11$^\circ$ were available if needed.  However, more recently
 the ISO-100 cross was replaced by a 30 cm dia. chamber to provide room for a moveable target system holding 4 gas targets \cite{gascell}, and 
 also the ability to place silicon (Si) detectors around the target for certain experiments.  In this configuration, the angular acceptance 
 of the system is between 2$^\circ$ and 5.25$^\circ$ in the lab.  For most experiments, the primary target is a 2.5 cm long gas cell containing ${^2}$H or ${^3}$He gas at atmospheric pressure (though solid targets are occasionally used).  The gas-cell windows
 are typically 4 $\mu$m Ti foils.  A list of the many publications carried out with {\it TwinSol} during its nearly 25 years of existence can
 be found here: \url {http://notredame.box.com/s/9w79ctrizr3ol8ac1pow9bckm5sa73bu}.
 
 \section{{\it TriSol}} 
 
 A further recent change in the setup shown in Fig.\ref{fig:1} was the removal of the neutron shielding wall shown and extension of the 
 beam line through a 1.25 m thick high-density concrete shielding wall into an adjacent area.  This modification provided additional isolation from
neutron and $\gamma$-ray background coming from the primary target and more room to attach ancillary equipment such as the $\beta$-decay station described in Ref.\cite{Halflife}.  Unfortunately, at this location the ion optics resulted in a
beam spot size of 25 mm FWHM or greater.  While acceptable for some experiments, this was problematic for others which resulted in
the first impetus to design an upgrade which would produce improved RNBs.  Additionally, the ``St. Benedict" ion trap \cite{StBenedict} requires a dedicated beam line.  Hence, reaction studies with exotic beams or other experiments including additional lifetime measurements could only be continued if two lines were
available.  Therefore, a small XY steering magnet followed by a magnetic dipole were inserted immediately after the shielding wall.
The dipole, which was originally part of the ORRUBA beamline \cite{ORNL} at Oak Ridge National Laboratory (ORNL), has ports at 0$^\circ$ and 15$^\circ$ to the {\it TwinSol} beam line.  The latter was used for {\it TriSol} to make use of its magnetic dispersion as discussed below.  Since the expected radioactive beams were much larger in extent than those used at ORNL, the clear aperture of this dipole was increased to 12 cm by the insertion of soft iron sections into its yoke.  Afterwards, the magnetic field was extensively mapped as a function of position and current, including the fringe field out to 65 cm.  The maximum central field of the modified magnet easily exceeded requirements for the most rigid beams that could be produced at {\it TriSol}.  The third superconducting solenoid \cite{Cryo1} has a clear, warm bore of 15 cm dia., a maximum central field of 5T, and a length of 1 m.  As was the case with the original two solenoids, the magnetic field was measured by the vendor and a file of the results was supplied with the magnet.  Its location was determined according to its clear aperture and the desired overall magnification of the system.  The 100 liter LHe supply of this solenoid is maintained within a recondensing cryostat \cite{Cryo2} by a cryocooler \cite{Cryo3}.  

 \subsection{Parameter calculation} 
 
Since the {\it TriSol} ion optics are more complex than those of {\it TwinSol}, it was necessary to use a simulation package such as LISE$^{++}$ \cite{LISE} \url{https://lise.nscl.msu.edu/lise.html} from MSU to design the spectrometer and enable calculation of its parameters.   A number of important operational considerations for this package are outlined in this section:
\\
\\  
(1) LISE$^{++}$ currently has no ability to directly simulate a gas target.  As a result, the energy, energy dispersion, and angular 
dispersion of the incident primary beam after the entrance window of the gas cell (typically a 4 $\mu$m Ti foil) must be calculated 
using the embedded ``Physical calculator" module, then entered into the ``Projectile" window.  The ``Stripper" is the exit foil.  The gas target between these two segments is treated as if it were a foil located at the center of the gas cell.  This introduces a small but not very serious error into the simulation.  For example, moving the target location by $\pm$1.25 cm (the length of the gas cell) results in a
$\pm4\%$ change in the calculated rate, a $\pm3\%$ change in the spot size, and a $\pm6\%$ change in the FWHM of the energy resolution.  The actual error will be of the order of one-half of these extreme values.
\\
\\
%\newpage
\noindent (2) The suggested method to determine the required B-fields for the magnetic elements is to scale the settings from a previous calculation according to the magnetic rigidity:
 
\begin{equation}\label{eq1}
R = B \rho = K\sqrt{\frac{ME}{Q^2}}.
\end{equation}

\noindent Here, M is the mass of the ion (generally taken to be the mass number A), E is its energy, Q is its charge state, $\rho$ is the radius of curvature of the ion's path, and K is a constant.  This will usually give initial settings within a few percent of the required values.  The final values can then be obtained by making small
changes in the parameters while observing the spot size and/or rate of the beam at various locations along the beam path.  
\\
\\
(3)  The production method is selected using ``Physics Models", three of which are relevant for {\it TriSol} calculations: (a) Fusion residuals, (b) Two-Body Reactions, and (c) ISOL mode.  Option (a) requires no additional input since the relevant cross sections and angular distributions are computed from an evaporation model.  Option (b) requires the input of cross sections, either directly from experimental data or from a reaction calculation using, e.g., a Distorted Wave Born Approximation (DWBA) code.  This is the most-used method.  The accuracy of the LISE$^{++}$ beam-rate predictions then depends on that of these cross sections, and will be best if there is experimental data at nearby energies.  The case of $^8$B, discussed below, is one of the more difficult ones.  The two-nucleon transfer process is calculated in a cluster model and there is no ``spectroscopic factor" to determine the absolute cross section.  Furthermore, there is only limited data at one energy far below that of interest to compare with.  The experimental yield was only about one third of the computed value.  However, the agreement with the calculation is much better for single-nucleon-transfer reactions.  Option (c) has only been used to for calibration purposes with an $\alpha$-particle source.  
\\
\\
(4) {\it TriSol} produces a ``cocktail" beam.  The most intense contaminant typically arises from scattering of the primary beam somewhere near the production target.  This component cannot be accurately predicted since the location of the source is distributed and its intensity is uncertain.  As a result, the contamination level of the beam must be experimentally determined.  Instructions on how to use LISE$^{++}$ to simulate {\it TriSol} beams, and some sample input files, are available at this \href {https://notredame.box.com/s/7s3rbyngwzwvc1pw5hydule02mzu03mz} {site}.  Comparisons between computed and experimental yields will be added as they become available.

 \subsection{Initial Results}

The beam line to St. Benedict is in the process of being completed up to the gas catcher, the first of its elements.  This is simply an
extension of the existing {\it TwinSol} line and so the beam transmission efficiency is well known, which has been confirmed by a
 LISE$^{++}$ calculation.  The calculated spot size is well within the acceptance of the gas catcher. 
  
\begin{figure}
[!htb]
\center
\includegraphics*[width=0.99\columnwidth]{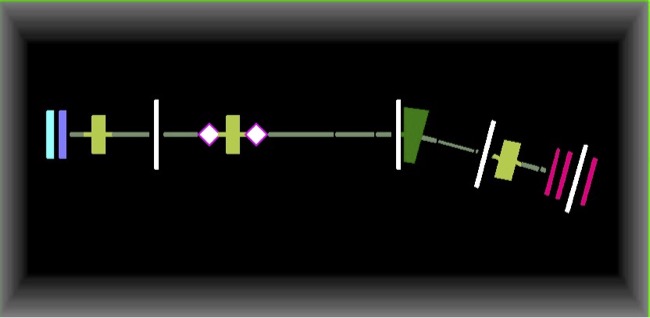}
\caption{\label{trisol}(Color online) {The \it TriSol} beam line.  The two bars on the left denote the primary target, the three solenoids are designated by green rectangles, and the dipole, located after the shielding wall, is illustrated by the dark-green wedge.  Collimators are shown as white lines and detector elements by red lines.  The two white diamonds are steering elements which do not physically exist but 
 are included in the simulation to test the influence of off-centered beam on the prediction.  In addition, two X-Y steerers (not shown) 
 are situated just before the dipole, and just after the dipole on the 15$^\circ$ line, to trim the beam position in the two lines as mentioned above.  (See Fig. \ref{fig:3} for the scale.)}
\label{fig:2} 
\end{figure}

\begin{figure}
[!h]
\center
\includegraphics*[width=0.99\columnwidth]{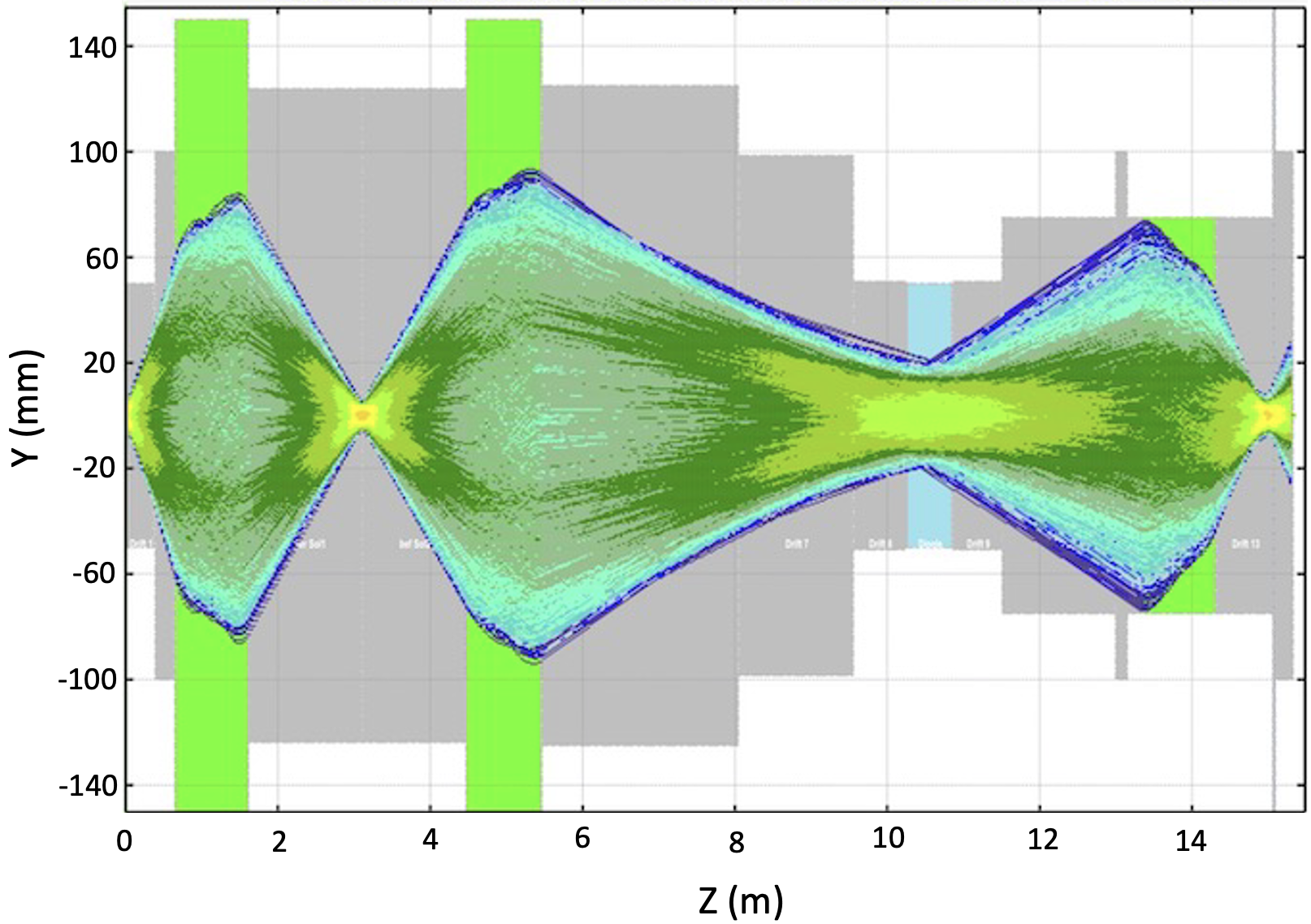}
\vspace{-3mm}
\caption{\label{mc}(Color online) The calculated $^8$B beam envelope.  This is a scale diagram showing the placement of the various elements of  {\it TriSol}.  The dipole is indicated by the blue rectangle centered at 10.5 m }
\label{fig:3} 
\end{figure}

 The {\it TriSol}  beam line has been completed.  A diagram of this line, taken from LISE$^{++}$ output, is shown in Fig. \ref{fig:2}.  A test run was carried out with a $^8$B beam to verify its performance.   The reaction was $^3$He($^6$Li, $^8$B) at an incident energy of 36 MeV, resulting in
 a 27.6 MeV $^8$B$^{5+}$ beam with a calculated intensity of 2.5x10$^4$ ions per second for 500 electrical nanoamperes (enA) 
 of $^6$Li.  The LISE$^{++}$ Monte-Carlo calculation of the beam envelope is illustrated in Fig. \ref{fig:3}. 

\begin{figure}
[!htb]
\center
\includegraphics*[width=0.99\columnwidth]{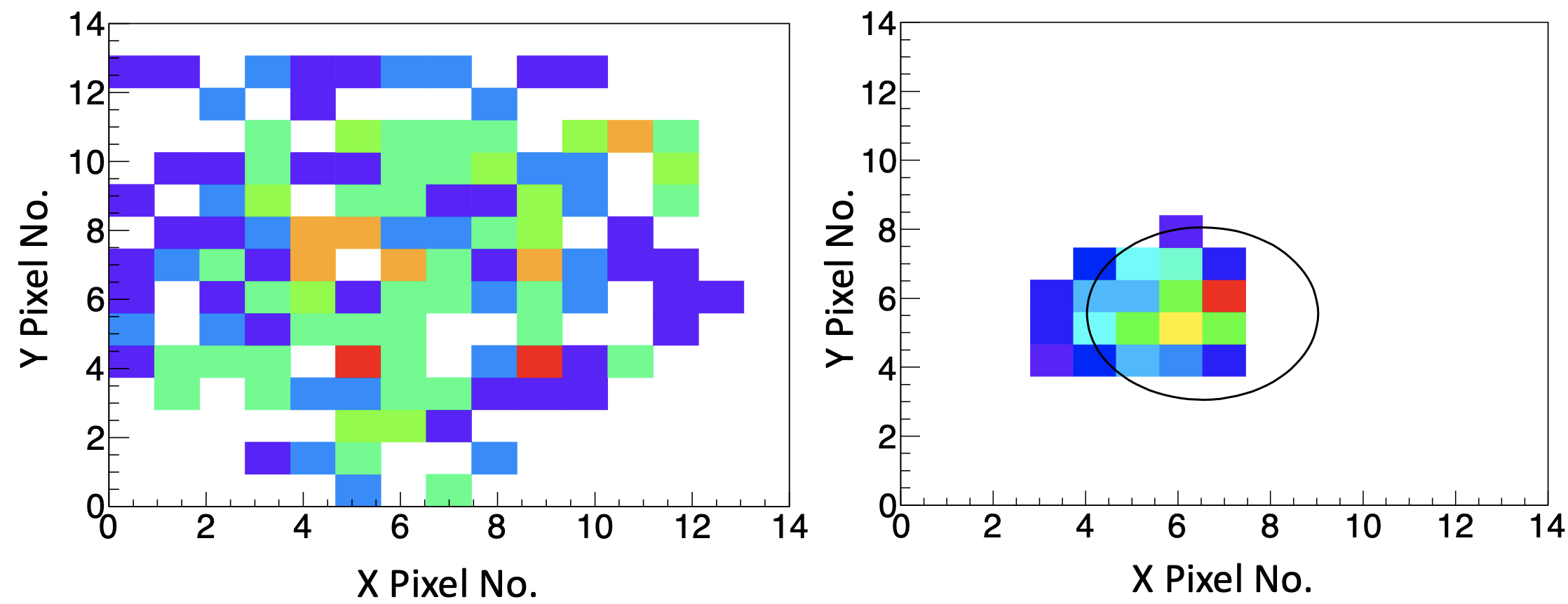}
\caption{\label{hp}(Color online) The measured hit patterns.  On the left is the pattern with the third solenoid turned off.
On the right is the hit pattern with this solenoid turned on.  These plots were taken for 965 $^8$B events on the detector with the
solenoid off and 913 events with the solenoid on.  The circle is the two-standard deviation diameter of the predicted hit pattern
for 913 events.}
\label{fig:4} 
\end{figure}

%\newline\noindent
The corresponding hit patterns at the focal point are shown in Fig. \ref{fig:4}, with and without current in the third solenoid.  The 
detector used was a 5 cm x 5 cm double-sided Si strip detector (DSSD) having a 3.3 mm x 3.3 mm pixel size.  In these images, the 
$^8$B beam is not exactly centered on the detector.  The solenoid-off hit pattern would otherwise extend beyond the left edge
of the DSSD.  The observed beam compression is compatible with the predictions of the Monte-Carlo simulation, thus verifying
the {\it TriSol} concept.  Referring to Fig. \ref {fig:3}, it can be seen that the divergence angle of the beam has been increased in
the process, as must happen due to conservation of the transverse emittance.  However, the maximum divergence is only 
$\pm$0.8$^\circ$, and there is sufficient room behind the solenoid to decrease this by at least a factor of two (at the expense of spot size) by moving the secondary target location downstream for those experiments that require smaller divergence.

\begin{figure}
[h!]
\center
\includegraphics*[width=0.90\columnwidth]{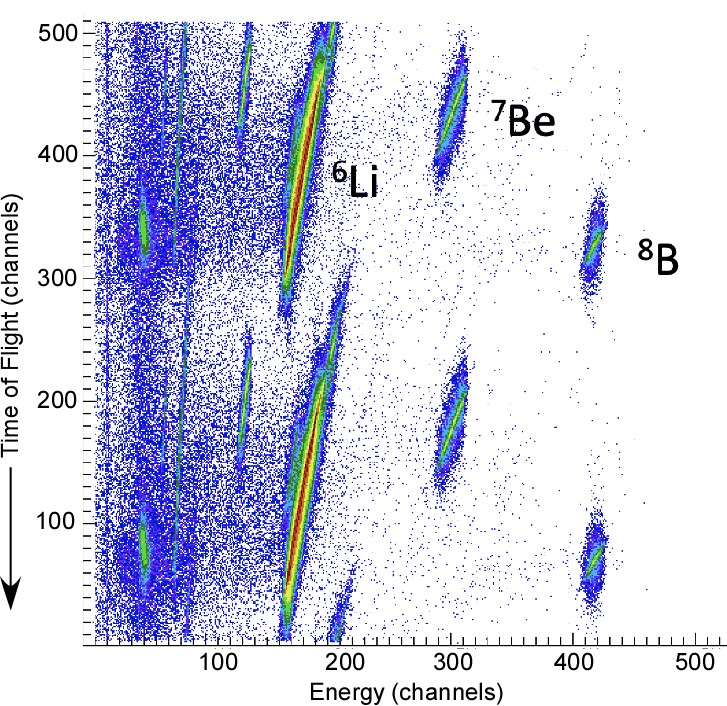}
\caption{\label{tof}(Color online) The timing spectrum from the $^8$B run;  time-of-flight on the Y-axis (0.39 ns/channel) vs. ion energy on the X-axis (66.2 keV/channel). 
For each ion, the charge state Q is equal to Z.}
\label{fig:5} 
\end{figure}

\begin{figure} [h!]
\center
\includegraphics*[width=0.95\columnwidth]{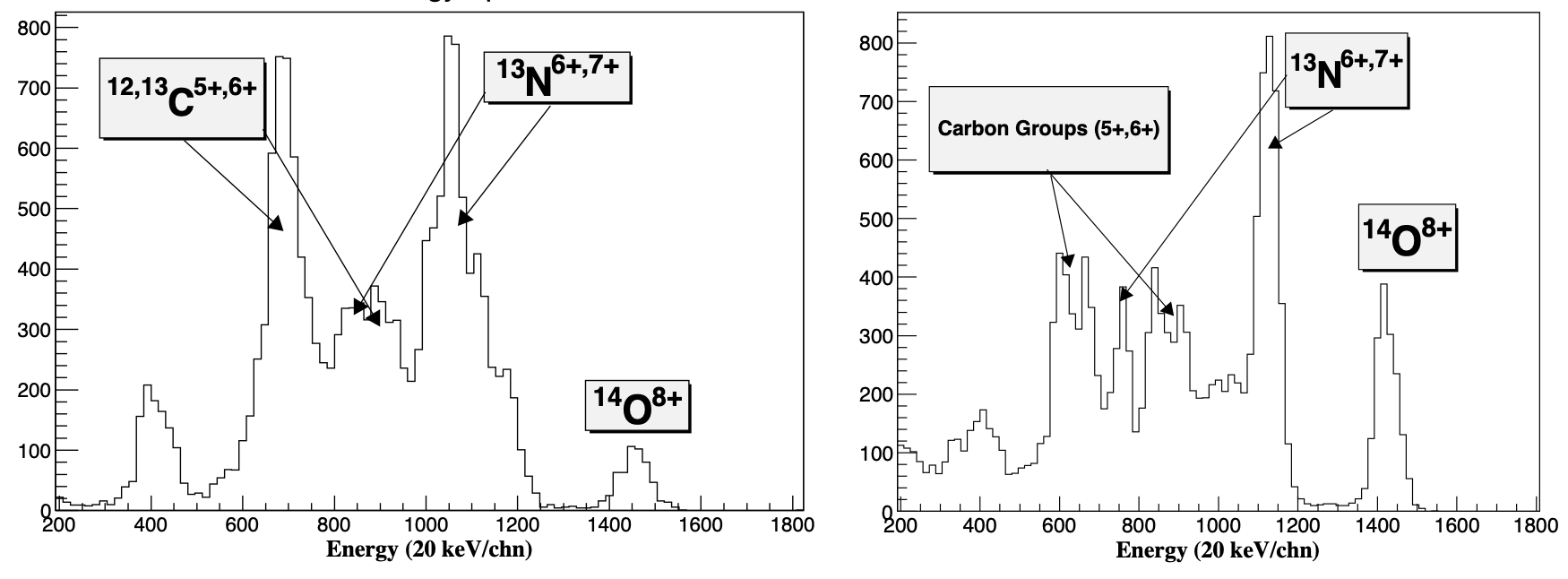}
\caption{\label{tof}Energy spectra for $^{14}$O production, without (left) and with (right) slits set to intercept the $^{12}$C contaminant.
The $^{12}$C peaks in the right spectrum are doubled due to scattering from two different locations.}
\label{fig:6} 
\end{figure}

Finally, several {\it TwinSol}  experiments (see, e.g., Ref. \cite{B8B}) required the use of time-of-flight (TOF) to select the relevant ion from a cocktail beam.
The Notre Dame FN tandem accelerator facility has the ability to produce a bunched beam with a time resolution of 2 ns (FWHM).  The resulting TOF spectrum for the $^8$B beam from {\it TriSol} is shown in Fig. \ref{fig:5}.  The main isotopes
\
%\newpage
%\noindent
in this spectrum are identified there.  The TOF is repeated twice due to the way the timing hardware operates.  As a result, the time difference between the two groups of similar ions, which is 101 ns, provides a calibration for this variable.  Observe that the TOF is highly correlated with energy so the ion's energy resolution can be improved using this feature, as discussed in Ref. \cite{TOF}.  In addition, note the strong $^6$Li groups from scattered primary beam particles.  In general, the TOF spectrum from {\it TriSol} is improved from that observed for {\it TwinSol} due to the longer flight path. 
%\newpage
\\
\\
Recently, we have been working to suppress contamination from scattered primary beam.  One method that seems to be effective in favorable cases involves charge-changing using a thin (400 $\mu$g/cm$^2$ Mylar) foil placed after the crossover collimator shown in Fig. 3.
In the case of $^{14}$O production that we have investigated, the yields of the 7$^+$ and 8$^+$ charge states at 30 MeV are comparable in intensity.   As a result, only a small amount of beam rate is lost when the first solenoid is set to pass 7$^+$ ions while the rest of the magnetic elements are set for the 8$^+$ charge state.  However, the contaminant $^{12}$C ions are focused into a ring around the $^{14}$O  ions immediately in front of the third solenoid.  A set of X-Y slits placed there can selectively pass $^{14}$O  while eliminating some of the $^{12}$C contaminant.  The reaction used was $^3$He($^{12}$C,$^{14}$O) at an incident energy of 54 MeV, with a $^{12}$C current of 1 e$\mu$A which is easily achievable.   In this case, there exists experimental data for the $^3$He reaction at an energy near the required value of 12.7 MeV \cite{14O}.  The result is an $^{14}$O beam energy of 29.5 MeV and a rate of 4.3x10$^4$/s (extrapolated from a measurement with a 1 enA primary beam current).  This is 85\% of the calculated rate when no slits are used.  However, the beam purity under these conditions is only 6\%.  The purity can be increased at the expense of beam rate by using the slits in front of the third solenoid to selectively intercept the $^{12}$C contaminant as discussed above.    Fig. \ref{fig:6} shows the result.  The fraction of $^{14}$O in the spectrum increased from 6\% to 18.5\%, but the rate was reduced by about a factor of three.  However, this is a much better option for active target detectors, such as the ND-Cube \cite{cube}, which have high efficiency but are rate limited due to pulse-pileup considerations.

%\newpage

 \section{Summary} 
 
 The {\it TriSol} project, nearing completion, provides separate beam lines for ion-trap experiments and nuclear reaction studies
 with radioactive nuclear beams.  The addition of a third solenoid, acting as a beam compressor, has resulted in beam spot sizes that are
 typically about 1 cm FWHM with a divergence angle of less than $\pm$1$^\circ$.  In addition, space exists behind the third
 solenoid to allow for some reduction of the divergence, at the expense of spot size, by moving the secondary target downstream.
 Time-of-flight measurement of the ions, often an important experimental parameter, has been improved due to the longer
 flight path of the system, and the location of the final focus, completely behind thick shielding walls, has essentially eliminated
 background from the primary target, thus improving the prospects for neutron and $\gamma$-ray experiments with radioactive beams.
 
The additional complexity of this system, compared with {\it TwinSol}, required the use of LISE$^{++}$ to simulate the spectrometer
 and thereby determine initial settings for the magnetic elements, which are typically within a few percent of their final values.  This also
 provided the ability to test new methods to improve the purity of the beam by reducing the scattered primary beam.  As an example,
 some {\it TwinSol} experiments introduced an absorber foil after the mid-plane collimator to purify the beam via differential energy
 loss.  However, LISE$^{++}$ simulations have shown that a thinner foil, which changes the charge of the ions between the solenoids, 
 is a far better purification method.  Under favorable circumstances, this can be accomplished with only a small reduction in the intensity of the desired beam.
 
With the completion of the {\it TriSol} project, coupled with an array of new instruments such as the ND-Cube active-target time projection chamber \cite{cube} and the St. Benedict ion trap, the faculty and staff of the Notre Dame Nuclear Science Laboratory, and its national and international collaborators, will be well-positioned to continue the ongoing studies of nuclear astrophysics, reaction mechanisms of exotic
 nuclear beams, and fundamental interactions.  
  
 %\newpage
 \section*{Acknowledgements}
 The {\it TriSol} project was supported by a grant from the University of Notre Dame, and by the US National Science Foundation
 under Grant $\#$2011890.  We also acknowledge important contributions from William von Seeger, Chevelle Boomershine, and
 Scott Carmichael, graduate students at Notre Dame, while co-author Sydney Coil  is an undergraduate student.  Professor Kolata
 especially wishes to thank Dr. Oleg Tarasov of Michigan State University for his help in implementing the LISE$^{++}$ simulation
 package for use with {\it TriSol}.
 
 \section*{Authorship credit contribution statement}
 
\textbf{P.D. O'Malley}: Supervision, Investigation, Writing $\&$ Editing, Visualization, Validation.  \textbf{T. Ahn}:  Investigation, Writing $\&$ Editing, Visualization.   \textbf{D.W. Bardayan}: Project Management, Investigation, Writing $\&$ Editing, Visualization.  \textbf{M. Brodeur}: Investigation, Writing $\&$ Editing, Visualization.  \textbf{S. Coil}: Magnetic dipole implementation, Magnetic field measurement.  \textbf{J.J. Kolata}: Conceptualization, Investigation, Writing - Initial draft, Editing, Visualization, Validation, Software implementation and validation.

\section*{Declaration of competing interest}

The authors declare that they have no known competing financial
interests or personal relationships that could have appeared to
influence the work reported in this manuscript.


\section*{References}

\begin{thebibliography}{1}

\bibitem{RIB1}
J.~J.~Kolata,~et~al., A radioactive beam facility using a large superconducting
  solenoid, NIMB 40/41 (1989) 503--506,
\newblock \href {http://dx.doi.org/10.1016/0168-583X(89)91032-X}
  {\path{doi:10.1016/0168-583X(89)91032-X}}.

\bibitem{TwinSol1}
Mu~Young~Lee~ (2002)~{\it TwinSol}: A dual superconducting solenoid ion-optical system for the production and study of low-energy radioactive nuclear beam reactions, Doctoral dissertation, University of Michigan, Ann Arbor, MI.

\bibitem{TwinSol2}
M.~Y.~Lee~et~al.,~{\it TwinSol}: A dual superconducting solenoid system for low-energy radioactive nuclear beam research, AIP Conf. Proc. 392 (1997) 397-400,
\newblock \href {http://dx.doi.org/10.1063/1.52712}
  {\path{doi:10.1063/1.52712}}.

\bibitem{TwinSol3}
F.D.~Becchetti, J.~Kolata, Low-energy radioactive beam experiments using the
  UM-UND solenoid RNB apparatus at the UND tandem: Past, present and future,
  AIP Conf. Proc. 392 (1997) 369-375,
\newblock \href {http://dx.doi.org/10.1063/1.52710}
  {\path{doi:10.1063/1.52710}}.
  

\bibitem{TwinSol4}
F.D.~Becchetti,~et~al., The {\it TwinSol} low-energy
  radioactive nuclear beam apparatus: status and recent results, NIMA 505~(2003)~377-380,
\newblock \href {http://dx.doi.org/10.1016/S0168-9002(03)01101-X}
  {\path{doi:10.1016/S0168-9002(03)01101-X}}.

\bibitem{TwinSol5}
P.D.~O'Malley, et al., Upgrades for the {\it TwinSol} facility, NIMB 376~(2016)~417-419,
\newblock \href {http://dx.doi.org/10.1016/j.nimb.2015.12.033}
  {\path{doi:10.1016/j.nimb.2015.12.033}}.
  
  \bibitem{TwinSol6}
F.D.~Becchetti,~et~al., Recent results from the {\it TwinSol} low-energy  RIB facility,
NIMB 376~(2016)~397-401,
\newblock \href {http://dx.doi.org/10.1016/j.nimb.2015.11.029}
  {\path{doi:10.1016/j.nimb.2015.11.2019}}.

\bibitem{pAT-TPC}
D.~Suzuki,~et~al., Prototype AT-TPC: Toward a new-generation active target time
  projection chamber for radioactive beam experiments, NIMA 691 (2012) 39-54,
\newblock \href {http://dx.doi.org/10.1016/j.nima.2012.06.050}
  {\path{doi:10.1016/j.nima.2012.06.050}}.

\bibitem{narray}
D.~W.~Bardayan,~et~al., Proton spectroscopic strengths of $^{18}$Ne,   
  AIP Conf. Proc. 2160 (2019), 070010,
\newblock \href {http://dx.doi.org/10.1063/1.5127733}
  {\path{doi:10.1063/1.5i2733}}.

\bibitem{gascell}
D.~W.~Bardayan,~et~al., Recent Nuclear Astrophysics Measurements using the TwinSol Separator,   
  J. Phys. Conf. Series 730 (2016), 012004,
\newblock \href {http://dx.doi.org/10.1088/1742-6596/730/1/012004}
  {\path{doi:10.1088/1742-6596/730/1/012004}}.

\bibitem{Halflife}
M.~Brodeur,~et~al., Precision half-life measurement of $^{17}$F, Phys. Rev. C 93 (2016) 025503,
\newblock \href {http://dx.doi.org/10.1103/PhysRevC.93.025503}
  {\path{doi:10.1103/PhysRevC.93.025503}}.
  
\bibitem{StBenedict}
M.~Brodeur,~et~al., V$_{ud}$ determination from light nuclide mirror transitions, NIMB 376 
  (2016) 376,
\newblock \href {http://dx.doi.org/10.1016/j.nimb.2015.12.038}
  {\path{doi:10.1016/j.nimb.2015.12.038}}.

\bibitem{ORNL}
S.~D.~Pain,~et~al., Development of a high solid-angle silicon detector array for measurement of transfer reactions in inverse kinematics, NIMB 261 (2007) 1122-1125,
\newblock \href {http://dx.doi.org/10.1016/j.nimb.2007.04.289}
  {\path{doi:10.1016/j.nimb.2007.04.289}}.
  
\bibitem{Cryo1}
Cryomagnetics, Inc. (Oak Ridge, TN)  050-591-HRTB-R.

\bibitem{Cryo2}
Cryomagnetics, Inc. (Oak Ridge, TN)  HRTB-591-R-408D2.

\bibitem{Cryo3}
Sumitomo Cryogenics of America, Inc. (Allentown, PA)  F70L.

\bibitem{LISE}
O.B.~Tarasov, D.~Bazin, LISE$^{++}$: Exotic beam production with fragment
separators and their design, NIMB 376 (2016) 185--187,
\newblock \href {http://dx.doi.org/10.1016/j.nimb.2016.03.021}
  {\path{doi:10.1016/j.nimb.2016.03.021}}.
  
\bibitem{B8B}
V.~Guimar\~aes,~et~al., Nuclear and Coulomb Interaction in $^8$B Breakup at  sub-Coulomb Energies, Phys. Rev. Lett. 84~(2000)~1862-1865,
\newblock \href {http://dx.doi.org/10.1103/PhysRevLett.84.1862}
  {\path{doi:10.1103/PhysRevLett.84.1862}}.

  
\bibitem{TOF}
F.D.~Becchetti,~et~al., Time-of-flight energy compensation to improve energy resolution in low-energy radioactive beam experiments at the {\it TwinSol} facility, NIMA 652~(2011)~532-536,
\newblock \href {http://dx.doi.org/10.1016/j.nima.2010.08.042}
  {\path{doi:10.1016/j.nima.2010.08.042}}.

\bibitem{cube}
T.~Ahn,~et~al., The Notre-Dame Cube: An active-target time-projection chamber for
radioactive beam experiments and detector development, NIMA 1025~(2022)~166180,
\newblock \href {http://dx.doi.org/10.1016/j.nima.2021.166180}
  {\path{doi:10.1016/j.nima.2021.166180}}.
  
  \bibitem{14O}
E.G.~Adelberger and A.B.~McDonald, A study of the $^{16}$O($^3$He,n) and $^{12}$C($^3$He,n) stripping reactions and a comparison of analogous ($^3$He,n), ($^3$He,p) and (t,p) transitions, Nuc. Phys. A 145~(1970)~497-533,

\newblock \href {http://doi.org/10.1016/0375-9474(70)90438-0}
  {\path{doi:10.1016/0375-9474(70)90438-0}}.





\end{thebibliography}
\end{document}